\documentclass[aps,prl,reprint,superscriptaddress,nobibnotes,showpacs]{revtex4-1}
\usepackage{graphicx}
\usepackage{multirow}
\usepackage{amsmath}
\usepackage{amssymb}
\usepackage{color}
\begin{document}

\setcounter{topnumber}{1}

\title{High spectral resolution of GaAs/AlAs phononic cavities by subharmonic resonant pump-probe excitation}

\author{Camille Lagoin}
\affiliation{Sorbonne Universit\'es, UPMC Univ.\ Paris 06, CNRS-UMR 7588, Institut des NanoSciences de Paris, F-75005, Paris, France}
\author{Bernard Perrin}
\affiliation{Sorbonne Universit\'es, UPMC Univ.\ Paris 06, CNRS-UMR 7588, Institut des NanoSciences de Paris, F-75005, Paris, France}
\author{Paola Atkinson}
\affiliation{Sorbonne Universit\'es, UPMC Univ.\ Paris 06, CNRS-UMR 7588, Institut des NanoSciences de Paris, F-75005, Paris, France}
\author{Daniel Garcia-Sanchez}
\email{daniel.garcia-sanchez@insp.upmc.fr}
\affiliation{Sorbonne Universit\'es, UPMC Univ.\ Paris 06, CNRS-UMR 7588, Institut des NanoSciences de Paris, F-75005, Paris, France}

\date{\today}
\begin{abstract}
We present here precise measurement of the resonance frequency, lifetime and shape of confined acoustic modes in the tens of GHz regime in GaAs/AlAs superlattice planar and micropillar cavities at low temperature ($\sim 20\,\textrm{K}$). The subharmonic resonant pump-probe technique, where the repetition rate of the pump laser is tuned to a subharmonic of the cavity resonance to maximize the amplitude of the acoustic resonance, in combination with a Sagnac interferometer technique for high sensitivity ($\sim 10 \,\textrm{fm}$) to the surface displacement, has been used. The cavity fundamental mode at $\sim 20\,\textrm{GHz}$ and the higher order cavity harmonics up to $\sim 180\,\textrm{GHz}$ have been clearly resolved. Mechanical Q-values up to $2.7 \times 10^4$ have been measured in a planar superlattice, and direct spatial mapping of confined acoustic modes in a superlattice cavity micropillar has been demonstrated. The Q-frequency product obtained is $ \sim 5 \times 10^{14}$ demonstrating the suitability of these superlattice cavities for optomechanical applications.
\end{abstract}


\maketitle

The field of optomechanics studies the interaction of an optical cavity with a mechanical resonator.
Recently optomechanical devices have attracted great interest for their potential in non-linear optics~\cite{SNaeiniNature2013} and for the demonstration of macroscopic quantum mechanics~\cite{RRiedingerNature2018,SHongScience2017,ASafaviNaeiniNature2013,WMarshallPRL2003}.
Superlattice micropillars have emerged as a promising choice for optomechanical experiments because they exhibit high acoustic resonance frequencies and large optomechanical coupling factors~\cite{AFainsteinPRL2013,DGarciaSanchezPRA2016} which are a requirement for non-linear optomechanics.
In addition superlattice micropillar cavities have been used to prepare single photon sources~\cite{NSomaschiNatPhot2016,XDingPRL2016} and to demonstrate polariton lasing~\cite{DBajoniPRL2008}.
The optical properties of the micropillars have been extensively studied; high quality factor optical cavities have been fabricated~\cite{CArnoldAPL2012,SReitzensteinAPL2007} and the optical modeshapes have been imaged~\cite{ctistis2010optical}.
Although preliminary theoretical~\cite{DGarciaSanchezPRA2016,lamberti2017optomechanical} and experimental~\cite{AFainsteinPRL2013,SAnguianoPRL2017} studies of the acoustic modes have been realized;
the full phononic characterization of confined modes in micropillars in the GHz regime poses significant experimental challenges, and standard optomechanical measurement techniques such as homodyne interferometry have not yet been successful.

Conversely the pump-probe technique is very well suited for the detection of high frequency 
acoustic phenomena, ranging from a few GHz up to $10\,\textrm{THz}$~\cite{perrin1999interferometric,GRozasPRL2009,huynh2006subterahertz,AFainsteinPRL2013,AAmzianePRB2011,TBienvilleUltrasonics2006}.
Recently this technique has been used to study micropillar cavities  with fundamental frequency  $\sim  19\,\textrm{GHz}$ and quality factor $\sim 1000$ at room temperature ~\cite{SAnguianoPRL2017,AFainsteinPRL2013}. 

In this work we have carried out pump probe measurements of the acoustic transmission coefficient of planar and micropillar GaAs/AlAs superlattice cavities at low temperature ($20\,\textrm{K}$).
In contrast to the room temperature measurements, where phonon-phonon interaction considerably reduces the mechanical quality factor, at low temperature the quality factor of the superlattice cavities is not limited by phonon absorption of the bulk material~\cite{RLegrandPRB2016}. An improvement of the phonon lifetime of crystalline materials such as GaAs by several orders of magnitude at low temperature has been observed~\cite{RLegrandPRB2016,MGoryachevAPL2012,SGalliouAPL2011,WChenPhysMagB1994}. 

However, the spectral resolution of standard pump probe experiments is limited by the repetition rate of the pulsed laser oscillator and/or the delay line and is insufficient to measure high intrinsic quality factors. We have overcome this frequency resolution limit by using a subharmonic excitation technique~\cite{ABruchhausenPRL2011} where the repetition rate of the laser pulse is tuned through resonance with a subharmonics of the acoustic mode.

\begin{figure}[t]
\begin{center}
\includegraphics{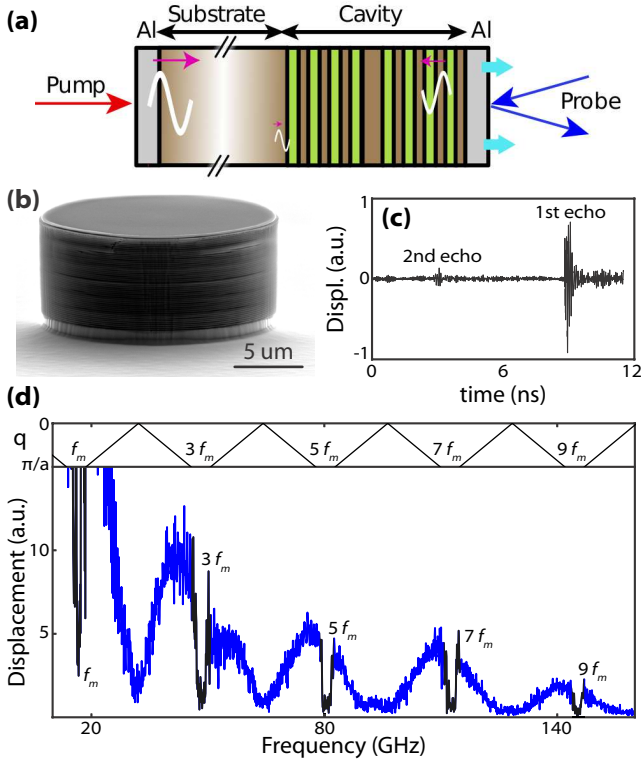}
\end{center}
\caption{
\textbf{(a)}
Device and measurement scheme.
\textbf{(b)} SEM picture of a $16\mu\textrm{m}$ diameter micropillar cavity. 
\textbf{(c)} Time trace of a $16\mu\textrm{m}$ diameter micropillar cavity obtained by pump-probe measurement.
Second echo arrives before first echo due to folding.
\textbf{(d)} Corresponding acoustic spectrum.
The schematic band structure can be found in the top part of the figure.
The bandgaps around the 1st, 3rd, 5th, 7th and 9th harmonics can be seen in the measurement.
}
\label{fig:SampleScheme}
\end{figure}

The technique presented here is a passive characterisation of the acoustic cavity. 
Opaque $60\,\textrm{nm}$ Al films deposited on the bottom and top surfaces of the samples, [see Fig.~\ref{fig:SampleScheme}(a)], provide an effective transfer mechanism for acoustic excitation and detection~\cite{huynh2006subterahertz} and suppress any photoelastic interaction between the optical beams and the superlattice cavity. In addition, the transmission configuration used here suppresses any cross-talk between the pump and probe.

A femtosecond pulse train excitation is generated by a mode-locked Sapphire laser with a tunable repetition rate $79.6 < f_\mathrm{rep}< 80.2\,\textrm{MHz}$ driven by a RF frequency generator. 
The pump pulse is focused on the backside of the sample which generates a broad-band acoustic pulse that propagates through the substrate and the acoustic superlattice cavity.
The pulse can be reflected several times by the backside and the frontside of the sample, giving rise to a series of echoes.
The travel time can be defined as the time that an echo takes to complete a round trip.
When the acoustic pulse reaches the frontside surface it generates a displacement that is probed using a Sagnac interferometer~\cite{EPeronnePRB2017} [see Fig.~\ref{fig:SampleScheme}(a)] which has higher sensitivity than the reflectometric technique in the $16-20\,\textrm{GHz}$ low frequency range~\cite{perrin1999interferometric}.
The two probes reach the sample surface with a time difference built to be equal to one half of the acoustic mode period $1/2f_\mathrm{m}$; this 
doubles the signal at the phononic resonance frequency.
The pump is modulated at $f_\mathrm{mod}\approx 1\,\textrm{MHz}$ and a lock-in amplifier is used to improve the signal-to-noise ratio.

The superlattice cavities discussed here are formed by two Distributed Bragg reflector (DBR) mirrors consisting of 15, 20 or 25 repeats of GaAs/AlAs bilayers either side of a $\lambda/2$ GaAs cavity that has been grown on a nominally $370\,\mu\textrm{m}$ thick GaAs (001) wafer by molecular beam epitaxy (MBE).
The micropillars have been etched by inductively coupled plasma etching [see Fig.~\ref{fig:SampleScheme}(b)].
This etching process results in very smooth and straight sidewalls~\cite{CArnoldAPL2012}.
The GaAs/AlAs layer thicknesses, and theoretically expected values of the cavity frequency, quality factor and amplitude lifetime of the three samples discussed here are given in table~\ref{tbl:samples}.

Fig.~\ref{fig:SampleScheme}(c) shows the time trace of the acoustic displacement, measured by the Sagnac interferometer, of a $16\,\mu\textrm{m}$ diameter micropillar from sample M20.
The second echo arrives at $3\,\textrm{ns}$ before the first echo ($9\,\textrm{ns}$) because of the time trace fold inside the measurement window.
This time arrival difference and the repetition rate can be used to determine the pulse travel time with great accuracy.

The Fourier transform of the time trace gives the corresponding acoustic spectrum [see Fig.~\ref{fig:SampleScheme}(d)].
Our pump-probe experimental setup allows all the phononic gaps of the superlattice from $16\,\textrm{GHz}$ up to $180\,\textrm{GHz}$ to be unequivocally identified.
These correspond to the theoretical Brillouin zone boundary gaps shown in the top part of Fig.~\ref{fig:SampleScheme}(d).
In our experiments, the spectral resolution is limited to $87\,\textrm{MHz}$ by the $11.5\,\textrm{ns}$ delay line.
As a result, our pump-probe technique, if the laser repetition rate is fixed, does not allow a resonance at $20\,\textrm{GHz}$ with a quality factor higher than 100 to be resolved.

\begin{table}[t]
\small
\centering
\begin{tabular}{|c|c|c|c|c|}
\hline
\multirow{2}{*}{name}&GaAs/AlAs&\multirow{2}{*}{$f_m^{th}\mathrm\,$(GHz)} &\multirow{2}{*}{$Q_\mathrm{th}$}&\multirow{2}{*}{$\tau_\mathrm{th}\,\textrm{(ns)}$}\\
&(nm)&&&\\
\hline
P15: planar&59.7/71.3 &$19.9$&$2.3\cdot 10^3$&$37$\\
\hline
M20: micropillars&75.1/88.7&$15.9$&$1.5\cdot 10^4$&$300$\\
\hline
P25: planar&60.0/70.1&$20.4$&$10^5$&$10^4$\\
\hline
\end{tabular}
\caption{Characteristics of the samples: thicknesses of the GaAs and AlAs mirror layers.
Expected theoretical properties: resonance frequency $f_m^{th}\mathrm\,$, quality factor $Q_\mathrm{th}$ and amplitude lifetime $\tau_\mathrm{th}$.}
\label{tbl:samples}
\end{table} 

To extract the quality factor by varying the laser repetition rate, we have to take into account the cumulative effect of the pump pulse train excitation~\cite{livrePerrin} in the pump-probe scheme.
The quadrature in-phase and out-of-phase outputs of the synchronous detection are given by $X(\tau)$ and $Y(\tau)$ where $\tau$ is the probe to pump delay time.
The complex output signals $s^\pm(\tau)=X(\tau) \pm i Y(\tau)$ are given by :
\begin{equation}
s^\pm(\tau)=\sum \limits^{+\infty}_{
p=0}\varphi(\tau+p/f_\mathrm{rep})e^{\pm i 2\pi  f_\mathrm{mod} (\tau+p/f_\mathrm{rep})}
\label{general_eq}
\end{equation}
where $ \varphi(\tau)$ is the displacement due to a single pump pulse.
When the frequency is close to the acoustic resonance of the cavity the displacement $\varphi(\tau)$ can be modeled by a damped harmonic oscillator motion $\varphi(\tau>0)\propto e^{-\tau/\tau_m} \cos( 2\pi f_m \tau)$, where $f_{m}$ is the acoustic resonance frequency and $\tau_m$ the acoustic amplitude lifetime.
Then equation~(\ref{general_eq}) becomes:
\begin{equation}
s^\pm(\tau)
\propto\frac{ e^{-\tau/\tau_m}e^{2i\pi f^\pm \tau} }{1-e^{-1/(\tau_m f_\mathrm{rep}) }e^{2i\pi f^\pm /f_\mathrm{rep}}}
\end{equation}
where $f^\pm \equiv(f_m \pm f_\mathrm{mod} )$.

The power spectrum $|{S}^\pm|^2(f)$ shows a maximum at $f^\pm$ corresponding to the cavity acoustic resonance.
The amplitude of $|{S}^\pm|^2(f^\pm)$ depends on the repetition rate and has a lorentzian shape centered around $f_\mathrm{rep}^n=(f_m \pm f_\mathrm{mod} )/n$ with $n$ a positive integer.
The subharmonic excitation is resonant when $f^\pm/f_\textrm{rep}=n$ and anti-resonant when $f^\pm/f_\textrm{rep}=n\pm1/2$.
The full width at half maximum~(FWHM) of the lorentzian resonance is given by $\Delta f=\frac{2 }{ n \tau_m}$ and can be used to calculate the lifetime $\tau_m$ and the quality factor $Q=\pi f_m \tau_m$ of the acoustic cavity.

\begin{figure}[t]
\begin{center}
\includegraphics{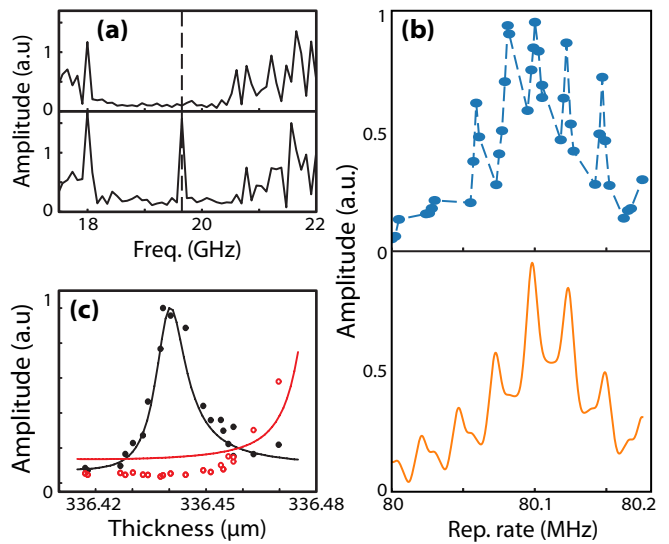}
\end{center}
\caption{
\textbf{(a)} 
Zoom in of the acoustic spectrum around the fundamental cavity mode.
The upper curve corresponds to a anti-resonant excitation with $f_\mathrm{rep}=79.990\,\textrm{MHz}$ and ${f^+}/{f_\mathrm{rep}}\simeq 245+1/2$.
The lower curve corresponds to a resonant excitation with $f_\mathrm{rep}=80.130\,\textrm{MHz}$ and ${f^+}/{f_\mathrm{rep}}\simeq 245$.
\textbf{(b)}
Acoustic resonance amplitude as a function of the laser repetition rate.
Measurement in upper curve and fit in lower curve.
\textbf{(c)}
Acoustic resonance amplitude as a function of the substrate thickness (proportional to the travel time).
The experimental data (dots) and fit (line) are shown for the two signals resonant at $f_m + f_{mod}$ in black and $f_m - f_{mod}$ in red.
}
\label{fig:Spectra}
\end{figure}

The lifetime of the acoustic resonance cavities that we have studied are larger than the laser repetition period of $\sim 1/80\mathrm{MHz}$. As a result, the acoustic resonance is not observed unless a specific repetition rate is chosen [see Fig.~\ref{fig:Spectra}(a)].

If the repetition rate is ${f_\mathrm{m}}/{f_\mathrm{rep}}\approx 245+1/2 $ the spectrum amplitude at resonance is completely suppressed because the pulse train excitation is anti-resonant with the phononic cavity (upper panel of Fig.~\ref{fig:Spectra}(a)).
On the other hand if the repetition rate is ${f_\mathrm{m}}/{f_\mathrm{rep}}\approx 245$, as in the lower panel of Fig.~\ref{fig:Spectra}(a) the resonance amplitude is at a maximum because the pulse train excitation is resonant with the phononic cavity: this is the subharmonic resonant excitation condition.

Fig.~\ref{fig:Spectra}(b) shows the amplitude of the cavity mode of the planar sample P15 versus the laser repetition rate.
A series of narrow peaks is superimposed over a broader peak.
This is due to the fact that the sample effectively consists of two coupled cavities: the superlattice cavity and the substrate which is itself an acoustic Fabry-Perot cavity.
The series of sharp peaks corresponds to the substrate cavity and the spacing between these peaks depends on the substrate thickness which is proportional to the echoes travel time (for this device $158\,\mathrm{ns}$).
The envelope of the peaks corresponds to the superlattice cavity resonance and the width is inversely proportional to the cavity quality factor.
From a simulation of the whole device (substrate and cavity), we can reproduce the variations of the resonance amplitude with the repetition rate [see orange line in Fig.~\ref{fig:Spectra}(b)] and estimate a cavity quality factor of $Q_\mathrm{meas}\simeq 1\,500$.

In order to test the high sensitivity of the subharmonic technique the laser repetition rate was set at the maximum of a sharp peak related to the substrate echo [see Fig.~\ref{fig:Spectra}(b)], and the sample was moved over a distance of approx. $0.5\,\textrm{mm}$ where we found a constant tickness gradient. 
The substrate thickness is not homogeneous across the wafer; therefore the echoes travel time depends on the position on the sample and as a result the series of sharp peaks from Fig.~\ref{fig:Spectra}(b) shifts in repetition rate.

The effect of varying the thickness of the substrate cavity on the vibrational displacement amplitude at a fixed repetition rate is shown in Fig.~\ref{fig:Spectra}(c). 
The two curves correspond to the two signals resonant at $f^\pm$. As the sharp echo peak width (1.3MHz) is smaller than twice the modulation frequency (3.2MHz), we can resolve the two resonances and observe that here the $f^-$ peak is maximal while the $f^+$ is minimized.

This shows that if the superlattice cavity resonance is broad, acoustic echos can have a significant effect on the acoustic mode response and care should be taken if measurements are carried out at fixed repetition rate at different positions on a sample.
To reduce the magnitude of the acoustic echoes we focus the pump beam more tightly on the sample backside to ensure that the acoustic wave diffracts more when propagating back and forth inside the sample.
The echo peaks also play less of a role when studying higher quality factor cavities and in small diameter pillars.

\begin{figure}[t]
\begin{center}
\includegraphics{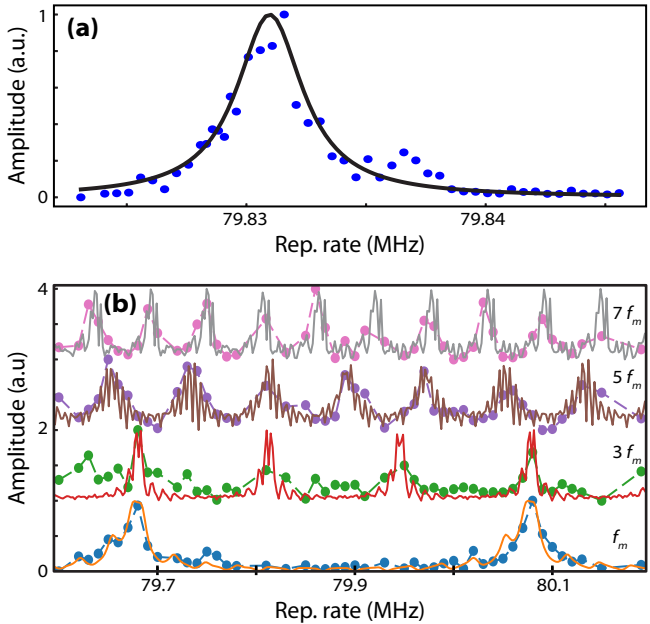}
\end{center}
\caption{
\textbf{(a)} Acoustic resonance amplitude as a function of the laser repetition rate for a 25 bilayer planar device.
\textbf{(b)}
Measured amplitude and fit of the fundamental acoustic mode of a $16\,\mu\textrm{m}$ diameter pillar, $f_\mathrm{m}$ (16.0GHz) and its odd harmonics, $3f_\mathrm{m}$ (48GHz), $5f_\mathrm{m}$ (80GHz), $7f_\mathrm{m}$ (112GHz).}
\label{fig:pillar}
\end{figure}

Fig.~\ref{fig:pillar}(a) shows the spectrum of a 25 bilayers DBR planar superlattice cavity from sample P25.
We measure the waist of the probe and pump optical beams which are respectively $2\,\mu\textrm{m}$ and $4.8\,\mu\textrm{m}$.
From the optical waist of the pump focalized on the backside, we can evaluate the waist of the acoustic excitation incident on the superlattice cavity taking into account the diffraction during the propagation inside the substrate and the GaAs anisotropy around the (100) axis. We find a waist of $14.3\,\mu\textrm{m}$ at $20\,\textrm{GHz}$.

The linewidth of the 25 repeat superlattice cavity is in the same order of magnitude than the free spectral range of the substrate cavity; as a result we do not observe the series of sharp peaks that we observed in Fig.~\ref{fig:Spectra}(b). 

This allows the quality factor of the superlattice cavity to be measured with good precision $Q_\mathrm{meas}=27\,200$ (corresponding to a lifetime $\tau_m=413\,\textrm{ns}$) which is approximately four times smaller than the theoretical quality factor $Q_\textrm{th}=10^5$.
At $20\,\textrm{K}$, the intrinsic damping of GaAs cannot explain this difference~\cite{RLegrandPRB2016}.
The reduction in Q is instead likely to be due to random thickness fluctuations of the MBE-grown DBR mirrors~\cite{GRozasPRL2009}. 

Fig.~\ref{fig:pillar}(b) shows the fundamental mode, the 3rd, the 5th and the 7th harmonics resonance amplitudes versus the repetition rate of a $16\,\mu\textrm{m}$ diameter micropillar from the sample M20.
The waist of the acoustic excitation at 16GHz incident on the superlattice cavity is estimated to be $17\,\mu\textrm{m}$.
Owing to the tunability of the repetition rate of our laser and its stability ($10\,\textrm{Hz}$), this technique allows the detection and study of acoustic resonances with frequencies in the range 10-$100\,\textrm{GHz}$ at least.
Indeed, in Fig.~\ref{fig:pillar}(b) for the fundamental mode curve, we can see two peaks which correspond to two different subharmonics resonant excitation and we observe as expected more subharmonic resonances for the higher order modes. We extracted a quality factor of $Q_\textrm{th}=7\,000$ ($\tau_m=135\,\textrm{ns}$) for the fundamental $16\,\textrm{GHz}$ mode which is roughly two times smaller than the theoretical quality factor again probably due to fluctuations in the mirror layer thicknesses.

Our technique also allows the mode profile of the micropillars to be investigated.
We scanned the micropillar surface by moving the probe beam along one direction of the micropillar. 
Figs.~\ref{fig:pillar}a-b shows the amplitude and the phase of the resonance of a $32\,\mu\textrm{m}$ diameter micropillar.
As expected for a longitudinal acoustic mode, the amplitude and the phase are axisymmetric.
We have also investigated the acoustic mode profile of a $64\,\mu\textrm{m}$ micropillar cavity (see Fig.~\ref{fig:pillar}c-d),
We observed several amplitude maxima which could be due to a hybrid mode and will be thoroughly studied in future works.

\begin{figure}[t]
\begin{center}
\includegraphics{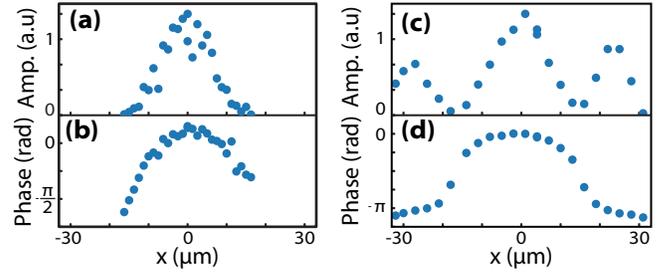}
\end{center}
\caption{Mode profile of a $32\,\mu\textrm{m}$ diameter micropillar: \textbf{(a)} amplitude and \textbf{(b)} phase.
Mode profile of a a $64\,\mu\textrm{m}$ diameter micropillar: \textbf{(c)} amplitude and \textbf{(d)} phase.}
\label{fig:mode_profile}
\end{figure}

In conclusion, we have demonstrated that the subharmonic resonant technique allows the frequency resolution limit of the standard pump probe measurement to be overcome and allows cavity mode lifetimes much larger than the time interval between two laser pulses to be measured.
We were able to measure a quality factor in the order of $2.7\cdot 10^4$ for a mode frequency as high as $20\,\textrm{GHz}$ which gives a $Q-f$ product of $5.0 \cdot 10^{14}$ at $20\,\textrm{K}$, comparable to state-of-the-art optomechanical photonic crystals~\cite{AGKrause2015PRL}.
It should be noted that the Q factors measured here were not limited by the measurement resolution and that the technique presented here should allow $Q-f$ products as large as $2\cdot 10^{16}$ to be measured.
In addition the measurement resolution permitted small shifts of $\sim 40\,\textrm{MHz}$ in the resonance frequency of different pillars to be clearly resolved - corresponding to sensitivity to small changes in the cavity thickness of $0.3\,\textrm{nm}$ due to a gradient in the growth thickness across the substrate.
Finally, we have demonstrated that this high resolution technique allows the mode profile of the acoustic mode to be imaged, opening up the possibility of testing theoretical models and tailoring acoustic modes to maximize the optomechanical coupling of superlattice micropillars in the future.

This work project was supported by the ``Agence Nationale de la Recherche'' and the ``Idex Sorbonne Universit\'es'' under contract No.\ ANR-11-IDEX-0004-02, through the MATISSE program.

\end{document}